\documentclass[article]{iopart}

\usepackage{graphicx}
\newcommand{\be}{\begin{equation}}
\newcommand{\ee}{\end{equation}}
\newcommand{\ba}{\begin{eqnarray}}
\newcommand{\ea}{\end{eqnarray}}
\newcommand{\ban}{\begin{eqnarray*}}
\newcommand{\ean}{\end{eqnarray*}}

\begin{document}

\title{Investigation of Hot QCD Matter:\\ Theoretical Aspects}
\author{Berndt M\"uller}
\address{Department of Physics, Duke University, Durham, NC 27708, USA\\
	Brookhaven National Laboratory, Upton, NY 11973, USA}

\begin{abstract}
This lecture presents an overview of the status of the investigation of the properties of the quark-gluon plasma using relativistic heavy ion collisions at the Relativistic Heavy Ion Collider (RHIC) and the Large Hadron Collider (LHC). It focuses on the insights that have been obtained by the comparison between experimental data from both facilities and theoretical calculations.
\end{abstract}

\maketitle

\section{Introduction}

The quark-gluon plasma produced in nuclear collisions at LHC and RHIC is a new form of matter with unique properties \cite{Jacak:2012dx,Muller:2012zq}: 
\begin{itemize}
\setlength{\itemsep}{0pt}
\item It is relativistic, yet strongly coupled;
\item it is a liquid that cools into a gas;
\item it is a nearly ``perfect'' liquid near the quantum limit of shear viscosity;
\item it thermalizes as fast as causality permits;
\item it creates its own new vacuum state to exist.
\end{itemize}
How is this possible?

Today we do not yet have a complete answer to this question, but theory has taken great strides towards developing frameworks in which it can be meaningfully addressed and experimental data can be used to clarify those properties of hot QCD matter that theory cannot yet reliably predict from QCD. The theory toolkit in relativistic heavy ion physics is quite diverse. It includes QCD perturbation theory in the vacuum and in a thermal medium (especially for the description of jets and heavy quarkonia); semiclassical gauge theory (for the description of the initial conditions reached in the nuclear collision); lattice gauge theory (for static thermodynamic properties of QCD matter, such as its equation of state and color screening); holographic methods mapping strongly coupled gauge theories on their gravity duals (for transport properties and the dynamics of thermalization); and transport theory, especially viscous hydrodynamics (for the evolution of the bulk matter).

\section{Properties of QCD Matter}

Lattice gauge theory has made impressive progress on the calculation of static thermodynamic properties of baryon symmetric QCD matter. The equation of state at $\mu_B = 0$ for physical quark masses is now known with a precision that exceeds that required in viscous hydrodynamics calculations \cite{Borsanyi:2010cj,Bazavov:2011nk,Borsanyi:2013bia}. The quasi-critical temperature where susceptibilities, related to chiral symmetry, peak has been determined to lie at $T_c \approx 155$ MeV. While resummed thermal perturbation theory \cite{Mogliacci:2013mca} describes most properties of the quark-gluon plasma well at temperatures above $2-3T_c$, its failure at lower temperatures indicates that QCD matter in the range $T_c \leq T \leq 2T_c$ is highly nonperturbative and strongly coupled, making its description theoretically challenging and interesting. 

\begin{figure}[htb]
\begin{center}
\includegraphics[width=4.5in]{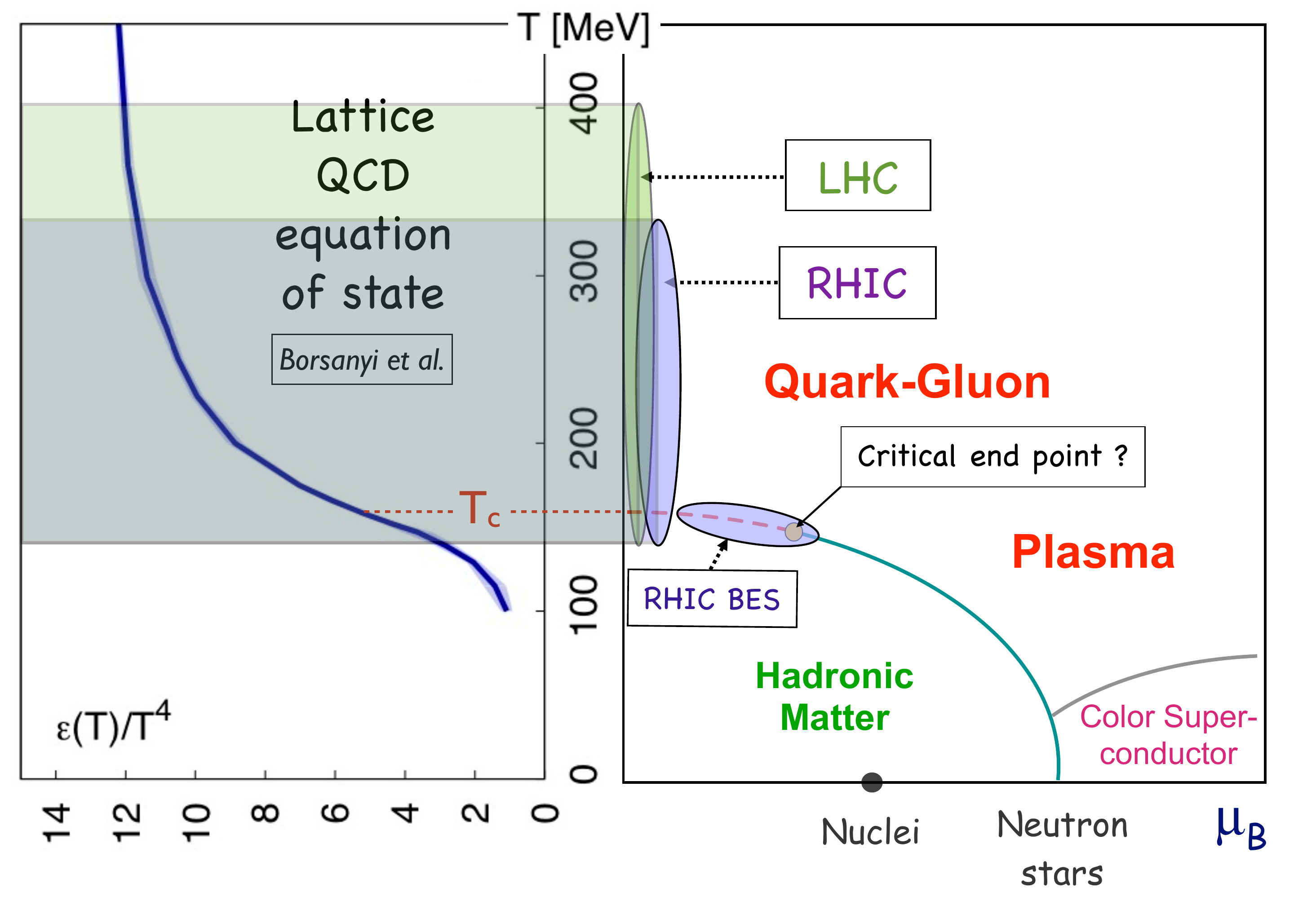}
\caption{Phase diagram of QCD matter (right panel) overlaid with regions covered by LHC and RHIC. The experimentally covered ranges are projected onto the energy density versus temperature at $\mu_B=0$ curve calculated by lattice QCD (left panel).\label{BM:fig1}}
\end{center}
\end{figure}

While RHIC can explore the temperature range only up to $2T_c$, the higher energy range of the LHC makes temperatures up to nearly $3T_c$ accessible (see Fig.~\ref{BM:fig1}). Experimental information about the temperatures reached in heavy ion collisions come from the measurement of the spectrum of radiated photons in the energy range $1-3$ GeV, where direct photon emission is dominated by thermal radiation. The measurement at a fixed beam energy only yields a temporal average of the temperature of the medium, but hydrodynamical simulations of the time evolution allow us to deduce a lower limit to the initial temperature. In Au+Au collisions at RHIC ($\sqrt{s_{\rm NN}} = 200$ GeV), this initial temperature exceeds 300 MeV \cite{Adare:2008ab,Adare:2009qk}; in Pb+Pb collisions at the LHC ($\sqrt{s_{\rm NN}} = 2.76$ TeV) the initial temperature is at least $30-40$\% higher \cite{Wilde:2012wc}. This increase is in good agreement with the observed scaling of the particle multiplicity from RHIC to LHC, which indicates an increase of at least a factor 3 in the initial energy density. 

\section{Probing QCD Matter}

There are many similarities between the evolution of the matter produced in a relativistic heavy ion collision (the ``little bang'') and the expansion of the matter filled early universe (the ``Big Bang''). In both expansions the initially imprinted quantum fluctuations propagate into macroscopic fluctuations in the final state via the acoustic and hydrodynamic response of the medium. In the Big Bang, the final temperature fluctuations probe the bulk dynamics; photons provide for penetrating probes, and light nuclei serve as chemical probes. In the little bang, the fluctuations in the final flow profile probe the expansion dynamics, photons and jets provide for the penetrating probes, and the various hadron species serve as the chemical probes. In each collision event the information that can be gathered from these probes is limited by the finite particle number; the advantage of the heavy ion experiments is that data can be gathered from many millions or even billions of collisions.

It is worthwhile asking which intrinsic properties of the quark-gluon plasma we can hope to determine experimentally and from which observables. A non-exhaustive list includes \cite{Muller:2012hr}:
\begin{itemize}
\setlength{\itemsep}{0pt}
\item 
The equation of state of the matter, given by relations among the components of the energy-momentum tensor $T_{\mu\nu}$ at equilibrium and their temperature dependence are reflected in the spectra of emitted particles. Lattice QCD is able to compute these quantities reliably.
\item
Transport coefficients of the quark-gluon plasma, especially the shear viscosity $\eta$, the coefficient $\hat{q}$ governing the transverse momentum diffusion of a fast parton (often called the {\em jet quenching parameter}), the coefficient of linear energy loss $\hat{e}$, and the diffusion coefficient $\kappa$ of a heavy quark, are related to the final-state flow pattern and the energy loss of fast partons that initiate jets. Lattice gauge theory presently cannot reliably calculate these dynamical quantities.
\item 
The static color screening length $\lambda_D$ (the inverse Debye mass $m_D$) governs the dissolution of bound states of heavy quarks in the quark-gluon plasma. This static quantity can be reliably calculated on the lattice.
\item
The electromagnetic response function of the quark-gluon plasma is reflected in the emission of thermal photons and lepton pairs. This dynamical quantity is difficult to calculate on the lattice, but moderate progress has been made recently.
\end{itemize}
All but the last of these properties is microscopically related to correlation functions of the gauge field. This implies that the associated experimental observables are mostly sensitive to the gluon structure of the quark-gluon plasma and only indirectly to its quark content. On the other hand, much more is known theoretically from lattice simulations about the quark structure of hot QCD matter, because it is much easier to construct operators from quark fields that can be reliably calculated. In this respect, lattice calculations and heavy ion experiments are to a certain degree complementary. The presence of jets in heavy ion collisions at LHC and RHIC tell us that at high virtuality $Q^2$ or high momenta $p$, the QGP is weakly coupled and has quasiparticle structure. On the other hand, the collective flow properties of the matter produced in the collisions tells us that at thermal momentum scales the quark-gluon plasma is strongly coupled. At which $Q^2$ or $p$ does the transition between strong and weak coupling occur? Does the quark-gluon plasma still contain quasiparticles at the thermal scale?  Which observables (jets?) can help us pinpoint where the transition occurs?

\begin{figure}[htb]
\begin{center}
\includegraphics[width=5in]{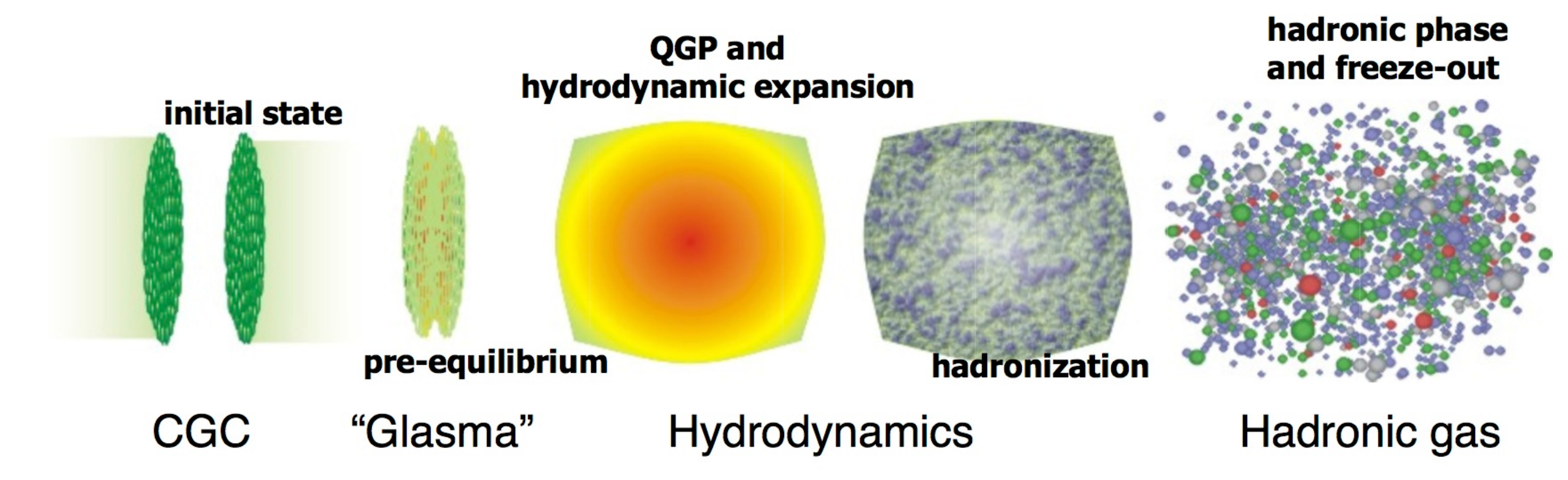}
\caption{Stages of a relativistic heavy ion collision in the ``standard model'' currently in use. The central stage of core interest is the quark-gluon plasma phase, which can be described by viscous hydrodynamics.\label{BM:fig2}}
\end{center}
\end{figure}

Maybe the most important theoretical achievement in the past decade is that a ``standard model'' of the dynamics of a relativistic heavy ion collision has emerged \cite{Heinz:2013wva,Hirano:2012kj,Fries:2010ht} (see Fig.~\ref{BM:fig2}). After a very brief period of equilibration -- most likely less than 1 fm/c -- the space-time evolution of the quark-gluon plasma can be described by relativistic viscous hydrodynamics with an exceptionally low shear viscosity-to-entropy density ratio. After cooling below $T_c$ the matter hadronizes. The subsequent expansion can be described by microscopic Boltzmann dynamics of a multi-component hadron gas until the gas becomes so dilute that all interactions cease and the hadron ensemble freezes out kinetically. The color glass condensate model of gluon saturation in the nuclear wave function at small $x$ has proven to be remarkably successful in predicting the initial energy and entropy deposition in the nuclear collision, and thus the initial conditions of the hydrodynamic evolution. This has made it possible to predict the bulk behavior over a wide range of collision energies and for a wide range of collision systems without arbitrary fit parameters. This remarkable achievement has been possible by the success of hydrodynamics during the dense phase of the collision where a microscopic description would be difficult to achieve owing to the strong coupling nature of the problem.

\section{The Perfect Liquid}

Hydrodynamics is the effective theory of the transport of energy and momentum in matter on long distance and time scales. In order to be applicable to the description of the quark-gluon plasma created in relativistic heavy ion collisions, which forms tiny, short-lived droplets of femtometer size, the hydrodynamic equations must be relativistic and include the effects of (shear) viscosity. The causal relativistic theory of viscous fluid has been worked out in full detail over the past few years \cite{Romatschke:2009im}. It is based on the framework of the M\"uller-Israel-Stewart formulation of second-order hydrodynamics, which includes relaxation effects for the dissipative part of the stress tensor. Schematically, the equations have the form
\begin{eqnarray}
\partial_\mu T^{\mu\nu} = 0 \qquad {\rm with} \qquad  T^{\mu\nu} &=& (\varepsilon +P) u^\mu u^\nu + \Pi^{\mu\nu} 
\\
\tau_\Pi (d\Pi^{\mu\nu}/d\tau) + \Pi^{\mu\nu} &=& \eta (\partial^\mu u^\nu + \partial^\nu u^\mu - {\rm trace}) .
\end{eqnarray}
It turns out that the quantity that most directly controls the behavior of the fluid, in addition to its equation of state, is the ratio of the shear viscosity $\eta$ to the entropy density $s$. The quantity $\eta/s$ is the relativistic generalization of the well known kinematic viscosity. Since in kinetic theory $\eta$ is proportional to the mean free path of particles in the fluid, which is inversely proportional to the transport cross section, unitarity limits how small $\eta$ can become under given conditions. An interesting consequence of this observation is that the quantity $\eta/s$ has a lower bound of the order of 0.08 (in units of $\hbar$). The existence of such a bound was conjectured almost 30 years ago \cite{Danielewicz:1984ww}, but it was quantitatively derived only recently using the technique of holographic gravity duals, the AdS/CFT duality \cite{Policastro:2001yc}. It is now believed that $\eta/s \ge (4\pi)^{-1}$ for any sensible quantum field theory \cite{Kovtun:2004de}. 

This so-called KSS bound is one aspect of a deep and fruitful relationship between the formation of black holes and thermalization in strongly coupled quantum field theories, which maps thermalization of a quantum field onto the process of formation of an event horizon in the dual gravity theory and recovers viscous hydrodynamiccs from the information absorbing dynamics of black hole horizons. Although a gravity dual for QCD is still unknown, the AdS/CFT duality has made it possible to explore thermalization and the approach to hydrodynamic behavior rigorously in strongly coupled gauge theories with some similarity to QCD \cite{CasalderreySolana:2011us}. This has put the models used to describe the dynamics of the quark-gluon plasma in relativistic heavy ion collisions on a much firmer basis.

\begin{figure}[htb]
\begin{center}
 \parbox{2.3in}{\includegraphics[width=2.2in]{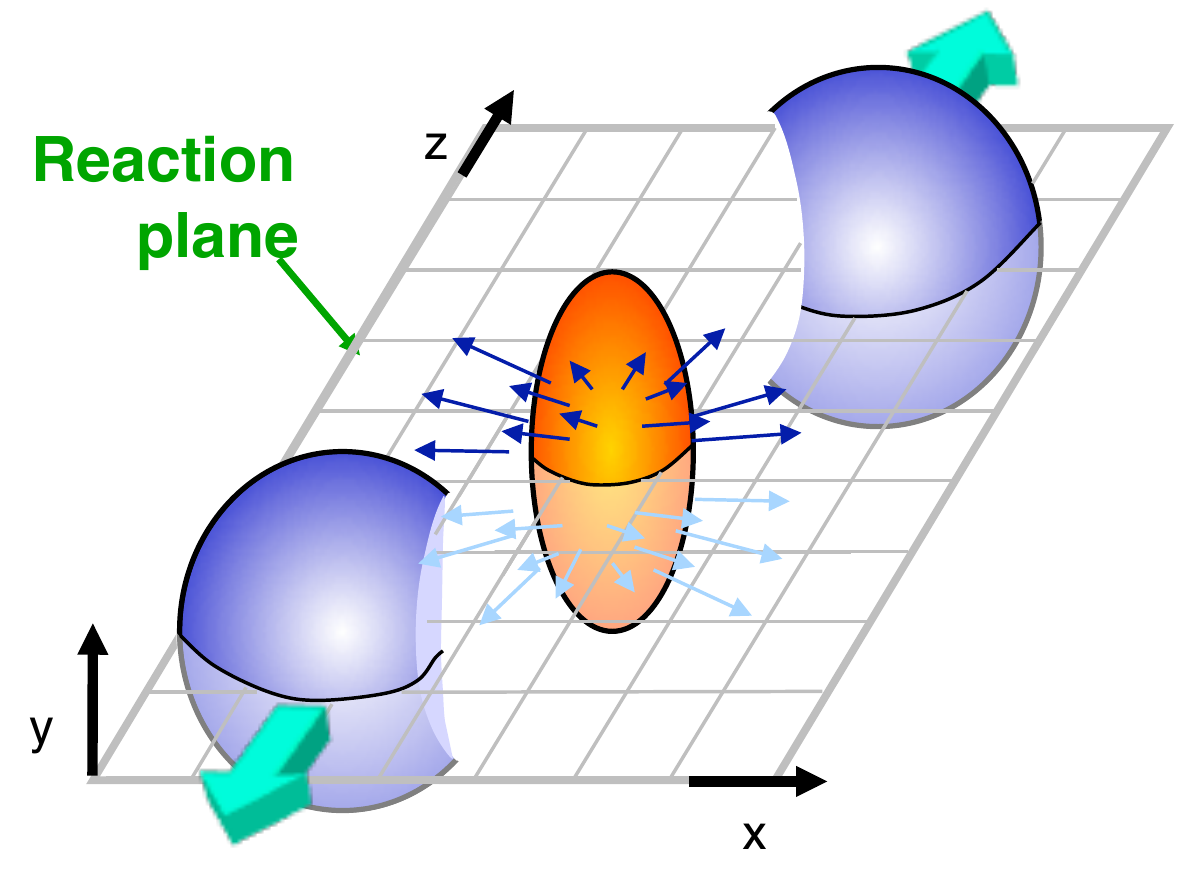}}
% \hspace*{0.2in}
 \parbox{2.4in}{\includegraphics[width=2.3in]{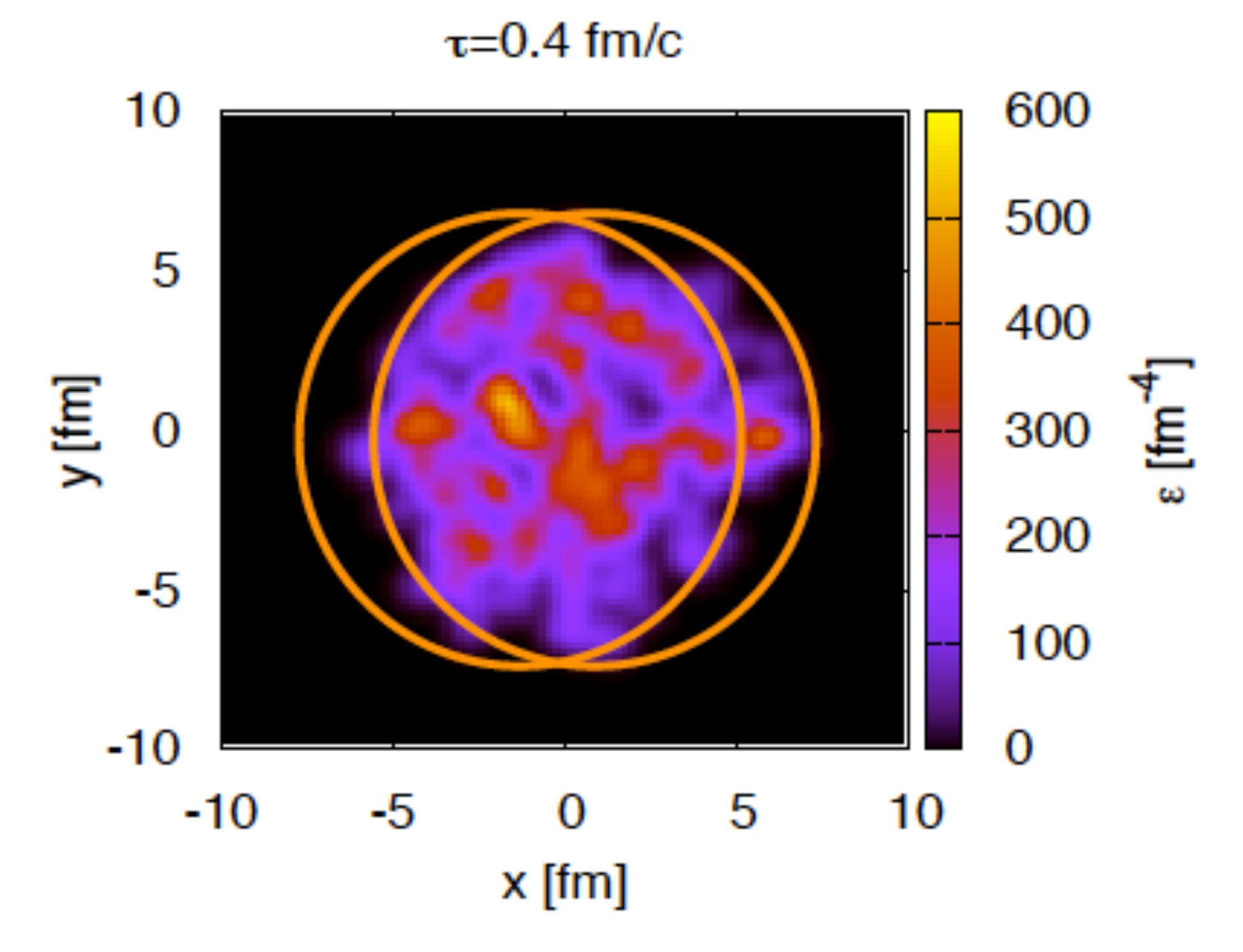}}
\caption{Geometry of a non-central heavy ion collision (left panel). Density fluctuations in the transverse plane in a sample collision event (right panel).\label{BM:fig3}}
\end{center}
\end{figure}

The experimental handle for the determination of $\eta/s$ is the azimuthal anisotropy of the flow of final-state particles in off-central heavy-ion collisions, where the nuclear overlap region is elongated in the direction perpendicular to the reaction plane, as shown in Fig.~\ref{BM:fig3}. Hydrodynamics converts the anisotropy of the pressure gradient into a flow anisotropy, which sensitively depends on the value of $\eta/s$ \cite{Song:2010mg}. The average geometric shape of the overlap region in symmetric nuclear collisions is dominated by the elliptic eccentricity, resulting in an elliptic flow anisotropy characterized by the second Fourier coefficient $v_2$. Event-by-event fluctuations of the density distribution within the overlap region generate higher Fourier coefficients for the initial geometry and final flow, encoded in higher Fourier coefficients $v_3$, $v_4$, etc. Their measurement is analogous to the mapping of the amplitudes of multipoles in the thermal fluctuations of the cosmic background radiation.

\begin{figure}[htb]
\begin{center}
 \parbox{2.2in}{\includegraphics[width=2.1in]{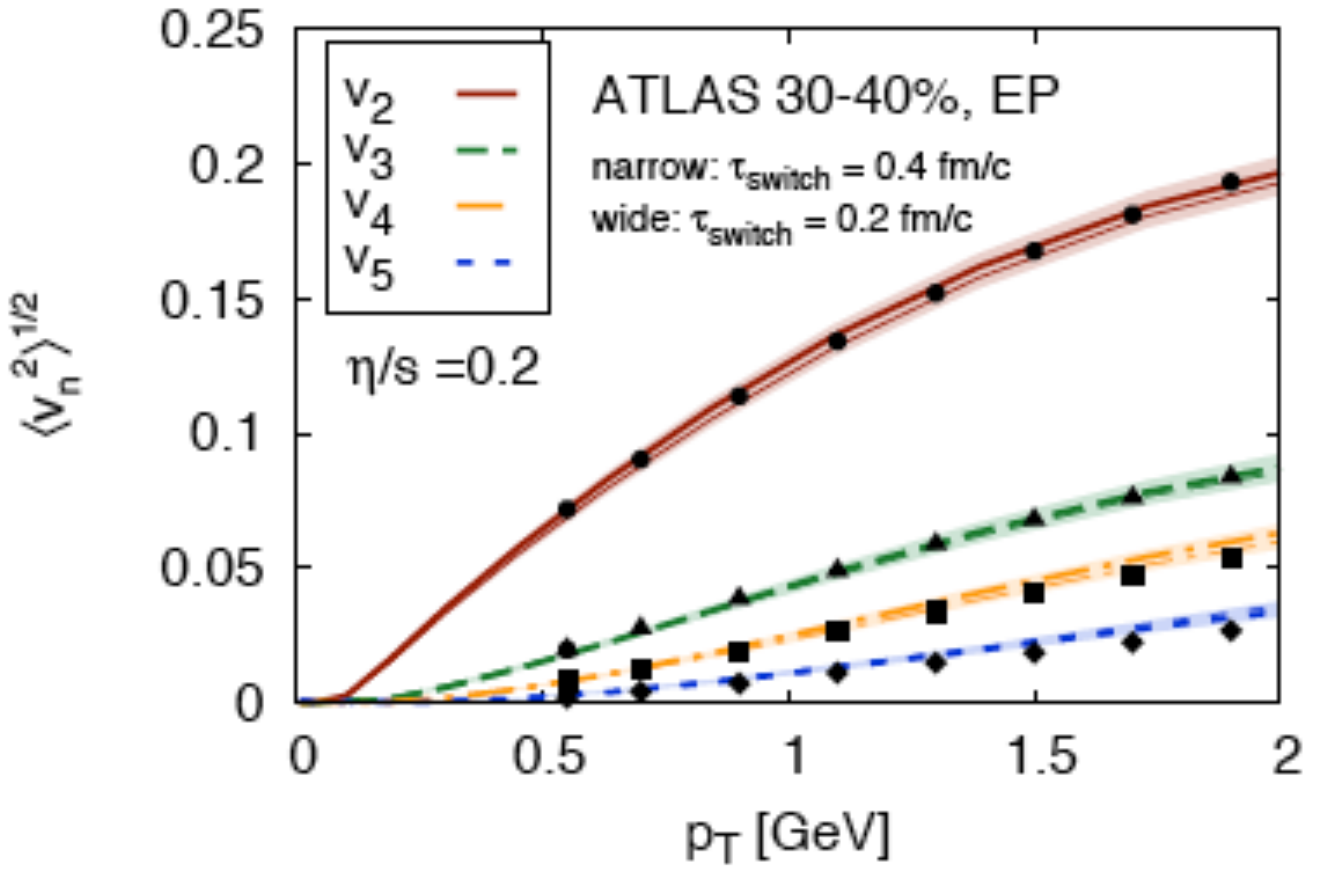}}
% \hspace*{0.2in}
 \parbox{2.6in}{\includegraphics[width=2.5in]{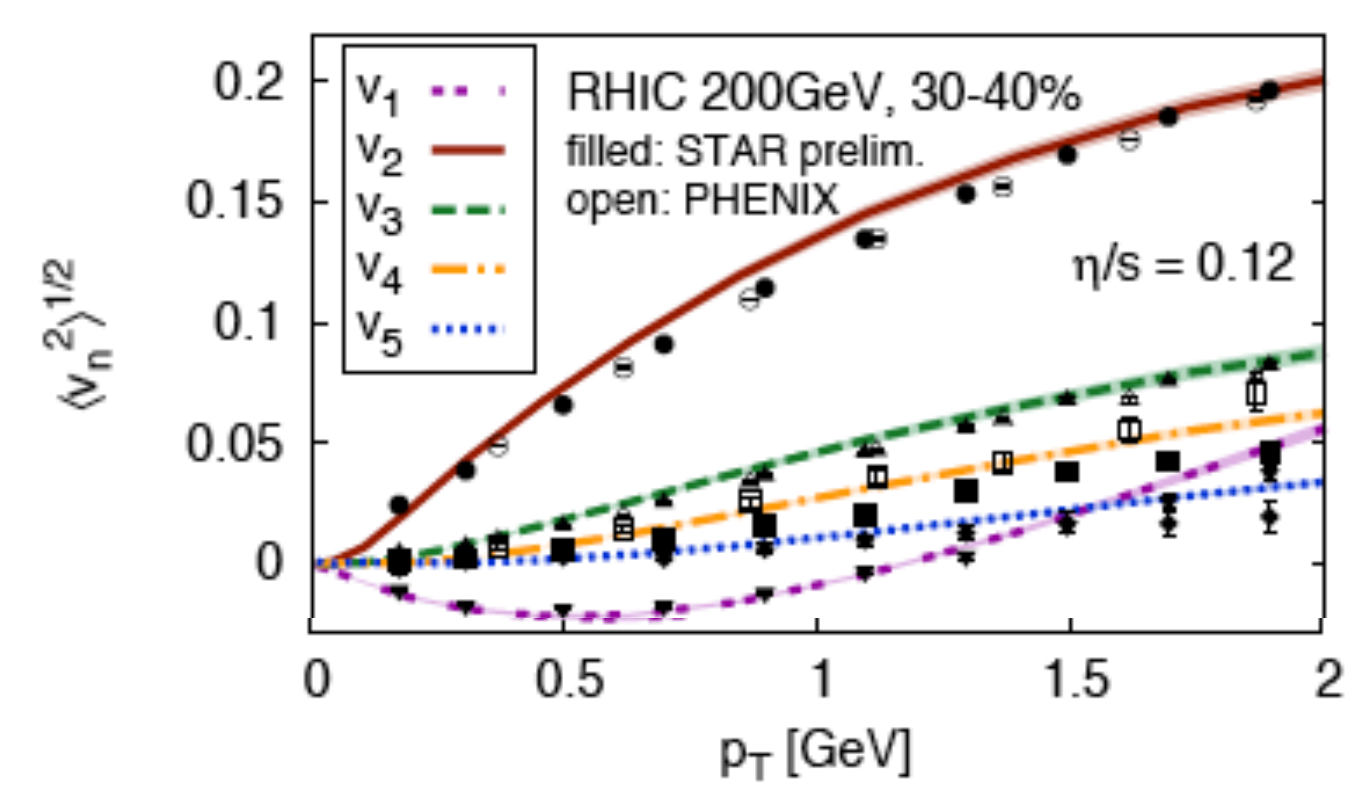}}
\caption{Fourier components of collective flow, $v_n(p_T)$, for Au+Au collisions at RHIC (left panel) and Pb+Pb collisions at LHC (right panel), in comparison with viscous hydrodynamics calculations. The deduced average value of the ratio $\eta/s$ is 60\% larger at LHC (0.20) than at RHIC (0.12).\label{BM:fig4}}
\end{center}
\end{figure}

The precise results of such an analysis of event-by-event fluctuations of the flow distribution depends somewhat on the structure of the initial-state density fluctuations, especially their radial profile and spatial scale. The most complete study of this kind to date \cite{Gale:2012rq,Gale:2013da}, starts from the fluctuations of the gluon distribution in the colliding nuclei, evolves them for a brief period using classical Yang-Mills equations, and then inserts the fluctuating energy density distribution into vicious hydrodynamics. The study concluded that the average value of $\eta/s$ (averaged over the thermal history of the expansion) in Au+Au collisions at the top RHIC energy is 0.12; whereas the value for Pb+Pb collisions at LHC is 0.20 (see Fig.~\ref{BM:fig4}). While each of these values has systematic uncertainties of at least 50\%, the ratio of these two values is probably rather stable against changes in the assumptions for the initial state. 

This analysis suggests that the average value of $\eta/s$ at the higher LHC energy is approximately 60\% higher than at RHIC \cite{Song:2011qa}, indicating a strong temperature dependence of this quantity (see \cite{Gale:2012rq}). It also indicates that the quark-gluon plasma at the lower temperature reached at RHIC is more strongly coupled and a more ``perfect'' liquid, making this energy domain especially interesting. Obviously, it would be of interest to make measurements of the flow fluctuations at energies between the top RHIC energy ($\sqrt{s_{\rm NN}} = 0.2$ TeV) and the present LHC energy ($\sqrt{s_{\rm NN}} = 2.76$ TeV). Future challenges include: Can we use p+A or d+A collisions to reduce the uncertainty of the initial state fluctuations in A+A collisions? Is the value for $\eta/s$ independent of the collision system (Cu+Cu, Cu+Au, U+U)? Do coherent color fields affect the early generation of flow, e.g.~in the form of an anomalous viscosity?

\section{Jet Quenching}

Energetic partons, the precursors of jets, lose energy while traversing the quark gluon plasma either by elastic collisions with the medium constituents or by gluon radiation \cite{Majumder:2010qh}. At high energies, radiation should dominate; collisional energy loss is expected to be important for intermediate energy partons and for heavy quarks. Each mechanism is encoded in a transport coefficient, $\hat{e}$ for collisional energy loss and $\hat{q}$ for radiative energy loss:
\begin{equation}
(dE/dx)_{\rm coll} = - C_2 \hat{e} , \qquad  (dE/dx)_{\rm rad} = - C_2 \hat{q} L ,
\end{equation}
where $L$ denotes the path length traversed in matter and $C_2$ is the quadratic Casimir of the fast parton. The value of $\hat{q}$ is given by the transverse momentum broadening of a fast light parton per unit path length.

\begin{figure}[htb]
\begin{center}
\includegraphics[width=5in]{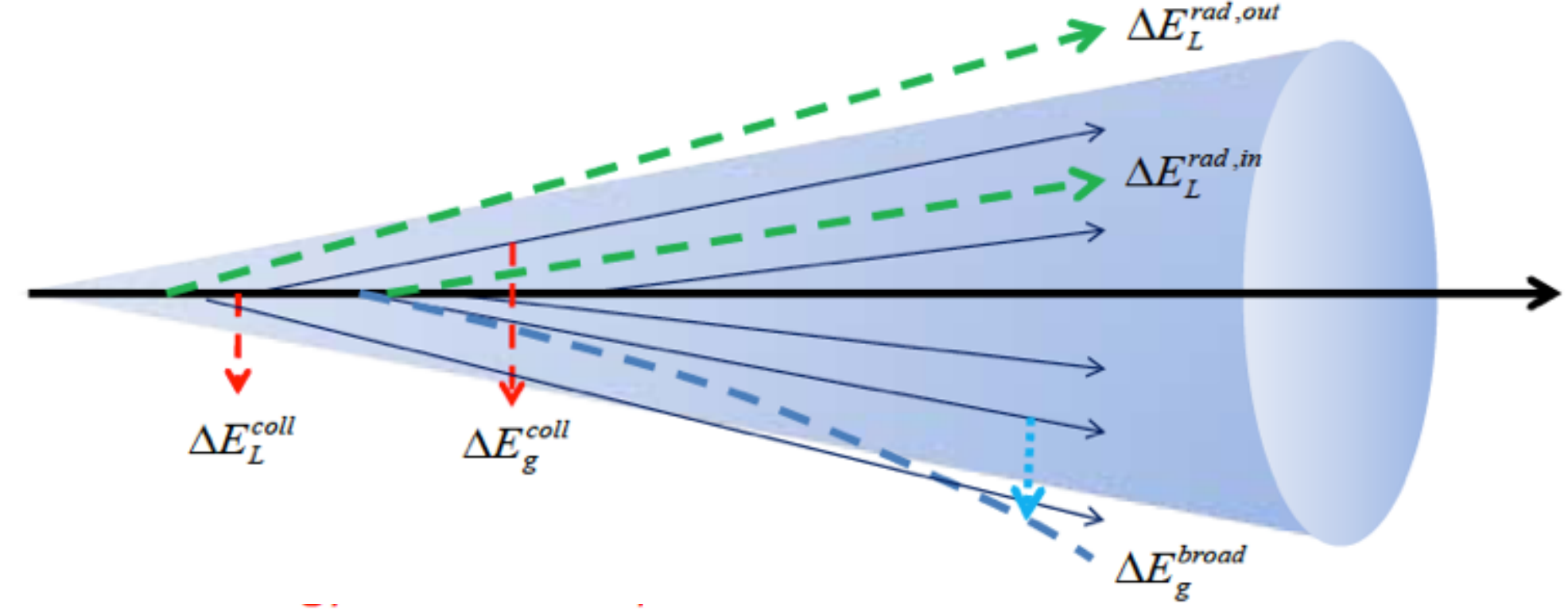}
\caption{Matter induced processes that contribute to jet quenching. The thick black arrow through the center of the cone shows the path of the primary parton. The thin black arrows indicate radiated gluons; the dashed green arrows represent soft radiated gluons that are deflected by the medium, some of them being kicked out of the jet cone. \label{BM:fig5}}
\end{center}
\end{figure}

The evolution of a jet in the medium, shown schematically in Fig.~\ref{BM:fig5}, is characterized by several scales \cite{Mehtar-Tani:2013pia}: The initial virtuality $Q_{\rm in}$ associated with the hard scattering process; the transverse scale at which the medium appears opaque, also called the saturation scale $Q_s$; and the transverse geometric extension of the jet, $r_\perp$. Those components of the jet, for which $r_\perp > Q_s^{-1}$, will be strongly modified by the medium. This implies that the core of the jet will lose energy coherently leading to an overall energy attenuation without modification of the shape of the jet core; but strong modifications are expected at larger angles and soft components of the jet. These features of jet modification can be encoded in a transport equation for the accompanying gluon radiation:
\begin{equation}
\frac{d}{dt}f(\omega,k_\perp^2,t) = \hat{e} \frac{\partial f}{\partial\omega} + \frac{\hat{q}}{2} \frac{\partial f}{\partial k_\perp^2}
+ \frac{dN_{\rm rad}}{d\omega dk_\perp^2 dt} ,
\end{equation}
where the last term denotes the gluon radiation induced by the medium.

The ``jet quenching parameter'' $\hat{q}$ can be determined by analyzing the suppression of leading hadrons in A+A collisions, compared with the scaled p+p data, usually given by a suppression factor $R_{AA}$, which is of the order of 0.2 for hadrons of transverse momenta in the range of 10 MeV/c in Au+Au at RHIC and Pb+Pb at LHC. Analysis of RHIC and LHC data show that $\hat{q}$ grows slightly less than linear (of order 10\%) with the matter density between RHIC and LHC \cite{Betz:2012fy}. This again indicates that the quark-gluon plasma formed at higher temperatures is less strongly coupled \cite{Buzzatti:2012dy}. Considering that $T^3/\hat{q}$ should scale like $\eta/s$ \cite{Majumder:2007zh}, one finds a somewhat weaker energy density dependence for dissipative effects at momentum scales relevant to jet quenching than at the thermal scale relevant to viscous effects. 

\begin{figure}[htb]
\begin{center}
\includegraphics[width=4.5in]{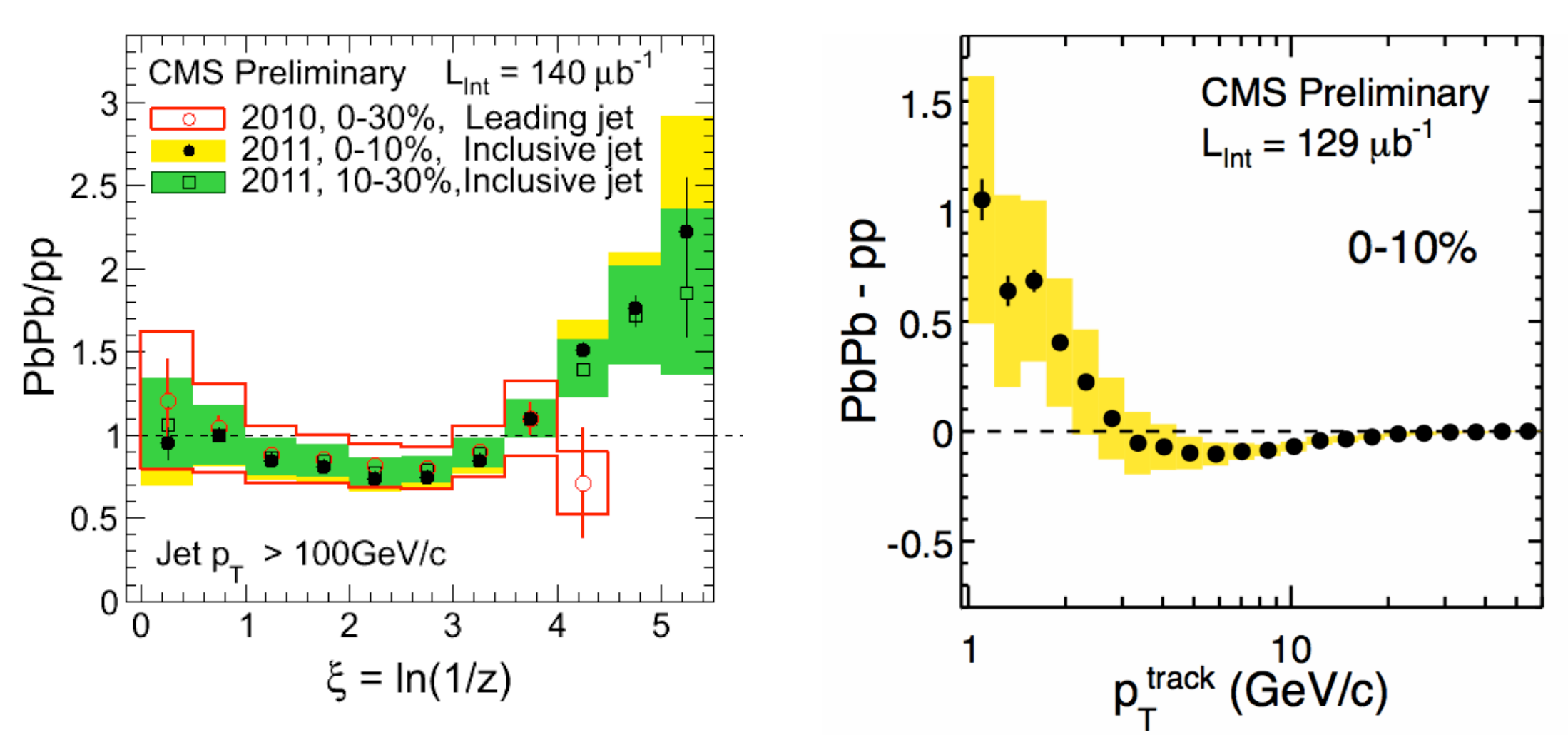}
\caption{Jet modifications observed in Pb+Pb collisions at LHC by the CMS collaboration. Left panel: Measured modification factor of the jet ``fragmentation function''. $z$ is the fractional energy carried by a hadron in the jet. Right panel: Measured relative change in the distribution of charged hadrons versus $p_T$ within jets.
\label{BM:fig6}}
\end{center}
\end{figure}

Using the values of $\hat{q}$ and $\hat{e}$ determined by comparison with $R_{AA}$ data, one can rather well explain the strong increase of the di-jet asymmetry observed in central Pb+Pb collisions at LHC \cite{Qin:2010mn}. This gives confidence that the basic mechanisms of jet modification and parton energy loss are reasonably well understood. The phenomenology of jet quenching at the LHC, exemplified by the CMS data from Pb+Pb collisions shown in Fig.~\ref{BM:fig6}, agrees qualitatively well with the expectation that modifications are concentrated at large cone angles and soft momentum fractions within the jet. A quantitatively reliable analysis awaits the creation of complete jet Monte-Carlo programs, which include the full panoply of radiative and collisional processes.

The LHC jet quenching data have clearly ruled out speculations that the jet could be strongly coupled to the medium. The perturbative theory of jet evolution works, with appropriate medium modifications, confirming that partons with energies of 10 GeV or more behave like ballistic quasiparticles inside the quark-gluon plasma \cite{Renk:2011wb,Zapp:2012ak}. The central remaining question is at which momentum scale partons become strongly coupled and lose their quasiparticle nature. This must be the case at thermal momenta (below 1 GeV) as evidenced by the ``perfect'' liquid behavior which is not compatible with a quasiparticle description of its constituents. Improved theoretical tools and precise resolved jet measurements at RHIC will be needed to identify the transition scale between weak and strong coupling.

\section{Quarkonium Melting}

\begin{figure}[htb]
\begin{center}
\includegraphics[width=4.5in]{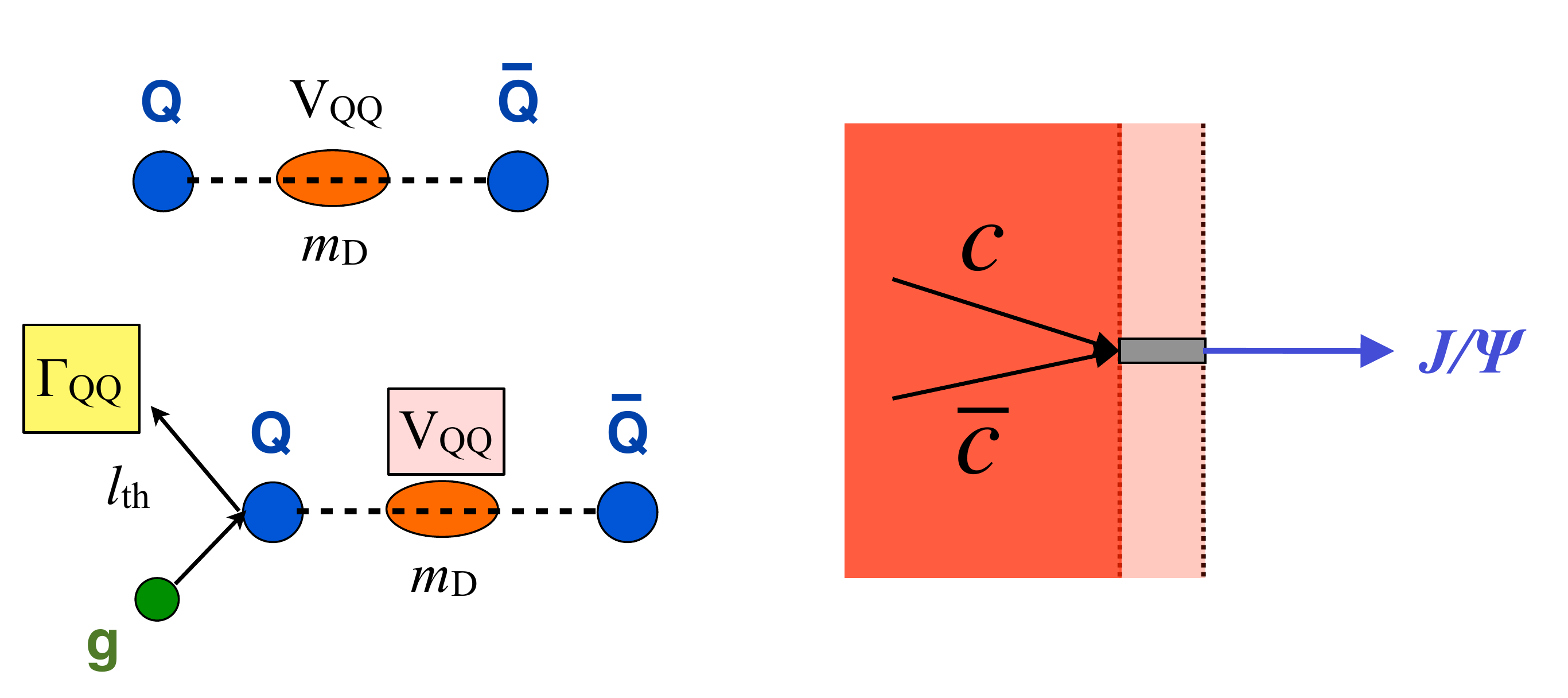}
\caption{Mechanisms contributing to matter induced changes in the yield of quarkonia: Color screening (upper left); ionization by thermal gluons (lower left); and recombination (right).\label{BM:fig7}}
\end{center}
\end{figure}

Bound states of heavy quarks, especially quarkonia ($J/\psi$, $\psi'$, the $\Upsilon$ states), are sensitive to the distance at which the color force is screened in the quark-gluon plasma. Several mechanisms contribute to nuclear modification of the quarkonium yield (see Fig.~\ref{BM:fig7}). At sufficiently high temperatures the screening length becomes shorter than the size of the quarkonium radius and the $Q\overline{Q}$ bound state ``melts''. Since the radii of the quarkonium states vary widely -- from approximately 0.1fm for the $\Upsilon$ ground state to almost 1 fm for $\psi'$ -- the sequential melting of these states could enable at least a semi-quantitative determination of the color screening length \cite{Karsch:1990wi,Karsch:2005nk}.  

The static screening length, which is relevant for heavy quarks, can be calculated on the lattice. However, it has become well understood in recent years that static color screening is only part of the picture of quarkonium melting, and that quarkonium yields can not only be suppressed by the action of the medium, but also enhanced by recombination, if the density of heavy quarks and antiquarks is large enough \cite{Rapp:2008tf}. An important loss mechanism is ionization by absorption of thermal gluons. This mechanism gains in importance as the binding energy of a quarkonium state is lowered by color screening \cite{Strickland:2011mw}. The absorption channel can be included in the dynamical evolution of the amplitude as an imaginary part of the potential with a corresponding noise term ensuring ultimate approach to the equilibrium distribution \cite{Akamatsu:2011se}:
\begin{equation}
i\hbar\frac{\partial}{\partial t} \Psi_{Q\overline{Q}} = \left[ \frac{p_Q^2 + p_{\overline{Q}}^2}{2M_Q} + V_{Q\overline{Q}} - \frac{i}{2}\Gamma_{Q\overline{Q}} + \xi_{Q\overline{Q}} \right] \Psi_{Q\overline{Q}} .
\end{equation}

\begin{figure}[htb]
\begin{center}
 \parbox{2.4in}{\includegraphics[width=2.3in]{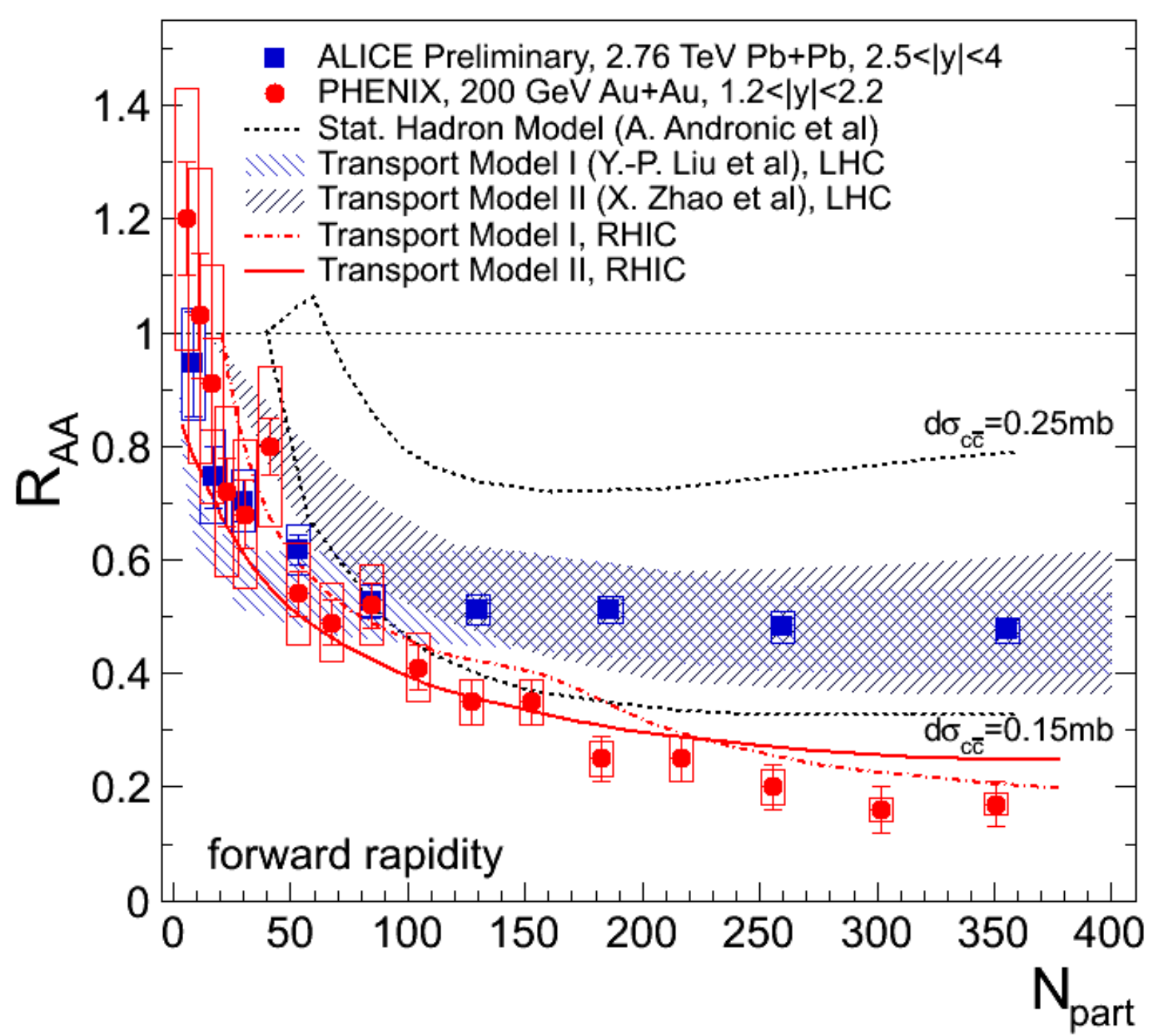}}
% \hspace*{0.2in}
 \parbox{2.4in}{\includegraphics[width=2.3in]{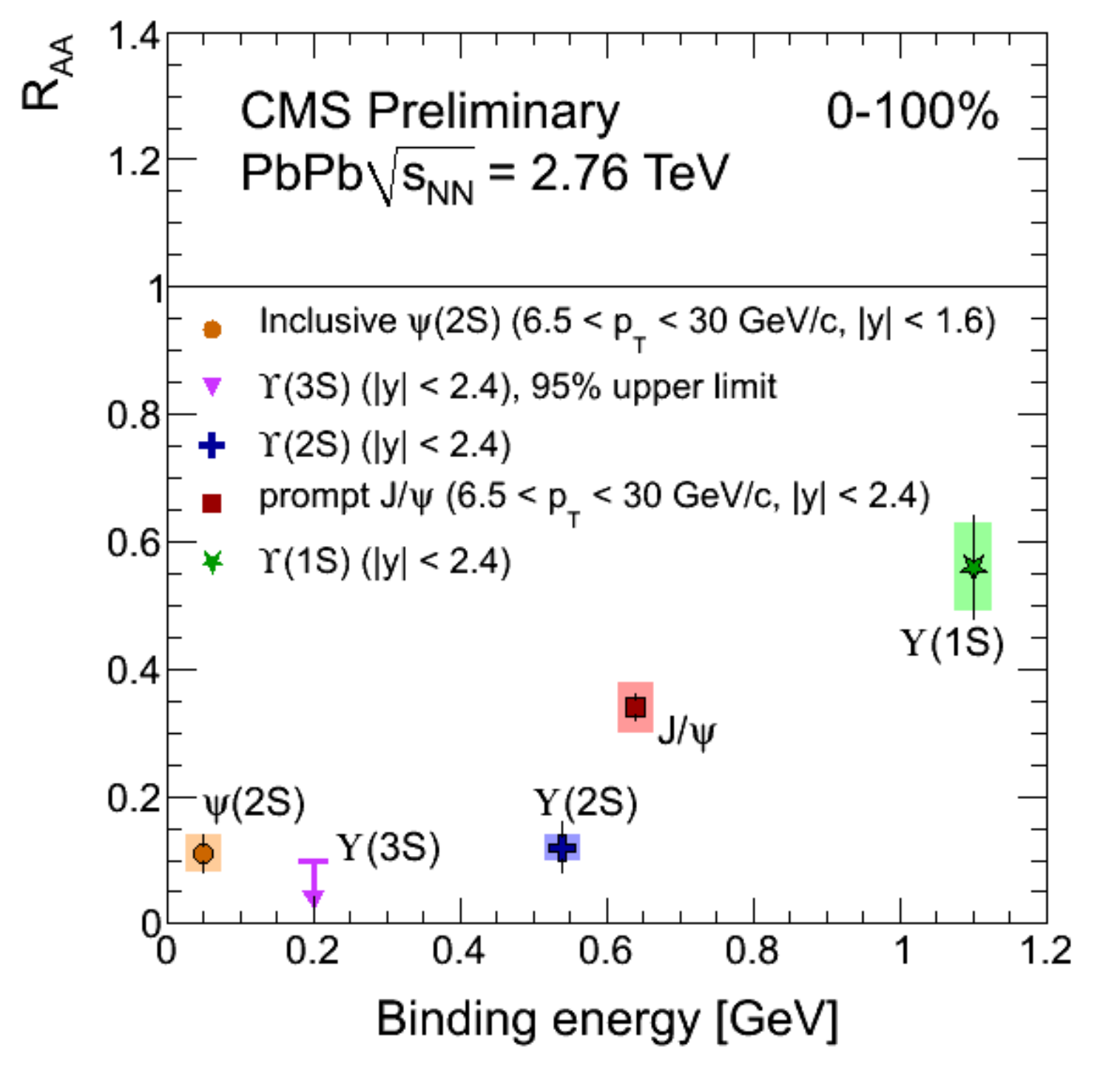}}
\caption{Left panel: Nuclear suppression factor $R_{AA}$ for $J/\psi$ as function of collision centrality in Au+Au at RHIC (red) and Pb+Pb at LHC (blue). Right panel: $R_{AA}$ in Pb+Pb at LHC for different quarkonium states. The surprisingly small amount of suppression for $J/\psi$ at LHC, compared with the suppression at RHIC (left panel) or with the suppression of the $\Upsilon$(2s) state at LHC (right panel)  is strong evidence for recombination.\label{BM:fig8}}
\end{center}
\end{figure}

Recombination of a heavy $Q\overline{Q}$-pair can occur at or near hadronization, similar to the sudden recombination mechanism that is thought to be responsible for the valence quark scaling of the identified particle elliptic flow. The yield of quarkonia formed in this manner grows quadratically with the heavy quark yield. Recombination of charm quark pairs into $J/\psi$ and $\psi'$ is thus expected to be much more frequent at LHC than at RHIC. This expectation seems to be borne out by a comparison of the centrality dependence of $J/\psi$ suppression observed by PHENIX in Au+Au at RHIC and by ALICE in Pb+Pb at LHC (see left panel in Fig.~\ref{BM:fig8}). The LHC data show {\em less} suppression in central collisions than the RHIC data, although the significantly hotter matter produced at LHC energy must surely be more effective in melting the $J/\psi$ state. The much weaker suppression at LHC (see right panel in Fig.~\ref{BM:fig8}) of the $J/\psi$ compared with the $\Upsilon$(2s), which has a similar binding energy and radius, is most readily explained by final-state recombination of $c$ quarks, which are much more abundant than the heavier $b$ quarks \cite{Zhao:2012gc}.  Whether it is possible to measure enough observables in order to not only disentangle the action of these different mechanisms, but also determine the color screening length, will need to be seen. On the positive side, the theory of quarkonium modification and transport in hot QCD matter has now reached a state of sophistication where this goal seems attainable if high statistics data can be measured at RHIC and LHC \cite{Mocsy:2013syh}.

\section{Summary}

The insights and open questions generated by the experimental results from LHC and RHIC can be summarized as follows:
\begin{itemize}
\item The quark-gluon plasma produced at the higher LHC energy is less strongly coupled than that produced at RHIC: The average value of $\eta/s$ at LHC is larger than at RHIC and the quark-gluon plasma at LHC appears less opaque than at RHIC.
\item Event-by-event fluctuations can be used as a versatile probe for the collective response of the quark-gluon plasma to differences in the initial conditions. The beam energy dependence can be used to vary the sensitivity to event-by-event fluctuations. There is reason for hope that details of the initial state structure can be separated from viscous effects, and both can be separately extracted from the data.
\item Jet physics opens new avenues of probing the quark-gluon plasma at different scales, because the matter effect on the jet structure creates probes of scales. An important goal is to find the kinematic threshold between quasiparticle and liquid domains in the quark-gluon plasma. The structure of the matter is reflected in the energy loss mechanisms and thus becomes amenable to investigation.
\item Quarkonium spectroscopy is blossoming with the advent of the LHC and its superbly suited detectors. It is now understood that quarkonium melting is not just a static screening phenomenon but involve other mechanisms, such as thermal ionization by gluon bombardment.  The first data from the LHC suggest that recombination plays a large role in central Pb+Pb collisions for the $c\overline{c}$ states.
\end{itemize}
Since our theoretical understanding of the formation and evolution of the quark-gluon plasma and its manifestation in observable phenomena is progressing rapidly and new experimental data are forthcoming at a torrid pace, there is justified hope that the remaining intriguing questions can be resolved in the years ahead.

\section*{Acknowledgments}

This work was supported in part by a research grant from the U.S. Department of Energy (DE-FG02-05ER41367).

\end{document}